\documentclass{article}

\usepackage{enumitem}
\usepackage{amsmath,bm} 
\usepackage{subcaption} 
\usepackage{multirow}
\usepackage{arxiv}
\usepackage{amsmath}
\usepackage[ruled,vlined]{algorithm2e}
\usepackage[utf8]{inputenc} 
\usepackage[T1]{fontenc}    
\usepackage{hyperref}       
\usepackage{url}            
\usepackage{booktabs}       
\usepackage{amsfonts}       
\usepackage{nicefrac}       
\usepackage{microtype}      
\usepackage{lipsum}		
\usepackage{graphicx}

\usepackage{doi}
\usepackage[utf8]{inputenc}
\usepackage[numbers]{natbib} 

\title{Transform and Bitstream Domain Image Classification}

\author{{P.R. Hill} \\
	Visual Information Laboratory\\ University of Bristol, UK\\
	\texttt{paul.hill@bristol.ac.uk} 
	\And
	{D.R. Bull}\\
	Visual Information Laboratory\\ University of Bristol, UK\\
	\texttt{dave.bull@bristol.ac.uk}  \\
}

\renewcommand{\shorttitle}{\textit{arXiv} Template}

\begin{document}
\maketitle

\begin{abstract}
Classification of images within the compressed domain offers significant benefits.  These benefits include reduced memory and computational requirements of a classification system.  This paper proposes two such methods as a proof of concept:  The first classifies within the JPEG image transform domain (i.e.\ DCT transform data); the second classifies the JPEG compressed binary bitstream directly.  These two methods are implemented using Residual Network CNNs and an adapted Vision Transformer.  Top-1 accuracy of approximately 70\% and 60\% were achieved using these methods respectively when classifying the Caltech C101 database. Although these results are significantly behind the state of the art for classification for this database (\~95\%), it illustrates the first time direct bitstream image classification has been achieved.  This work confirms that direct bitstream image classification is possible and could be utilised in a first pass database screening of a raw bitstream (within a wired or wireless network) or where computational, memory and bandwidth requirements are severely restricted.
\end{abstract}


\section{Introduction} 
Classification performance of standard image databases has undergone rapid improvements over the last decade.  Significantly improved results over previous methods (such as SIFT based methods \cite{lowe2004distinctive}) were initially achieved using deep Convolutional Neural Networks: CNNs (using the so called AlexNet CNN \cite{krizhevsky2012imagenet}).  Alexnet combined a conventional CNN architecture (proposed by LeCun \cite{lecun1998gradient}) with a large number of layers (at the time), more effective activation functions (ReLU \cite{glorot2011deep}) and the use of an extremely large database and the increased computational power of GPUs (compared to conventional used CPUs).  More recent innovations have further improved image classification performance.  These include:

\begin{description}[leftmargin=0cm, style=sameline]
\item [Spatial Pyramid Pooling (SPP): ]
 Spatial Pyramid Pooling was introduced in order to effectively manage input images of varying sizes and resolutions \cite{he2015spatial}.
\item [Inception: ] Inception\_v3 \cite{szegedy2016rethinking} was considered to give state of the art classification performance when it was created (as was its predecessor Inception\_v2).  These networks added a larger number of layers and added architectural features such as the factorisation of larger spatial convolutions.
\item [ResNets: ]  There are many types of residual networks (e.g.\ \cite{he2016deep}) and they have also given state of the art classification results.  Their main innovation was the inclusion of skip steps within network architectures.
\item [Vision Transformers (ViTs): ] ViTs have taken the concept of attention previously used for NLP applications \cite{vaswani2017attention} and extended it to image classification \cite{dosovitskiy2020image}. Currently, Scaling Vision Transformers hold the best performance for classifying the C101 database~\cite{zhai2021scaling}.
\end{description}
 
Conversely, many different image compression methods and standards have been used over the previous decades (JPEG \cite{wallace1992jpeg}, JPEG2000 \cite{rabbani2002jpeg2000}, HIFIC \cite{mentzer2020high} etc.).  It has been recognised that the transform domain of these compression systems show similarities to the transform domains used for image classification used before the era of deep learning and CNNs~\cite{hill2002wavelet}.  It is therefore not unreasonable to expect that classification within these transform domains could be as effective as directly using images for classification.  Furthermore, the advantage of using the transform (or bitstream) domain is that there is a potentially significant reduction in computational, bandwidth and memory requirements (a reflection of the high levels of compression achieved by image compression systems).

\subsection{Transform and Bitstream Domain Classification}

A number of authors have recently reported excellent results in reconstructing compressed images using trained neural networks~\cite{zhu2018image,souza2019hybrid,niu2020endtoend}. 
These sources have inspired the use of networks for classification rather than reconstruction.  Although JPEG can be considered in some cases to be an outdated image compression system, it was chosen as the target transform and bitstream classification domain within our work as it still has extremely high usage globally and is simple to manipulate, analyse and represent in our classification system.

True compressed domain classification, using a compressed bitstream as input to the classifier, is extremely challenging.  This is due to the decorrelated nature of the bitstream and the complex and interacting transform, prediction and entropy coding components that it comprises.  A small number of authors have attempted to achieve compressed domain classification \cite{cheheb2021video,alvar2018can, chadha2017compressed}.  However, these methods have all used video compressed bitstreams and decode only specific tokens (such as motion vectors or coding modes) for content characterisation.  No previous methods have directly used the actual bitstream for classification.

Due to the sparse and spatially heterogeneous structure of intra-coding processes employed in codecs such as HEVC and VVC, we instead adopted the uniformly tiled DCT (Discrete Cosine Transform) based structure (as used in JPEG) for initial analysis.  This spatial consistency is essential for effective analysis using standard 2D CNN layers.  

\begin{figure}[ht!]
\begin{center}
\includegraphics[]{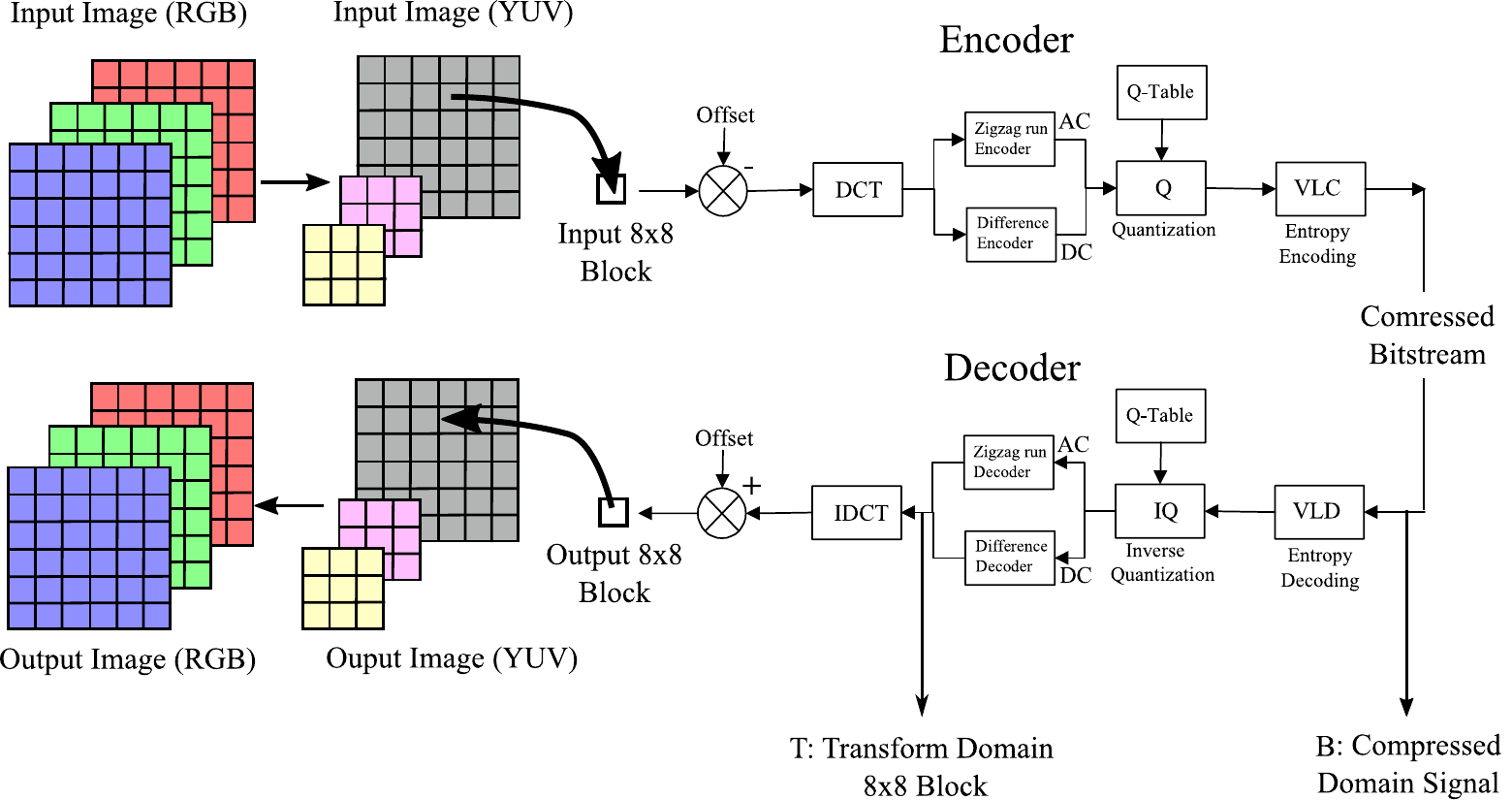}
\caption{Basic DCT based JPEG image codec architecture.  This figure shows the compressed ($B$) and transform domain ($T$) features that used to classify.}
\label{fig:jpeg1}
\end{center}

\end{figure}

Figure \ref{fig:jpeg1} shows the basic structure of a JPEG DCT based codec.  This shows an image transformed into the commonly used YUV (4:2:0) format and subsequently spatially tiled into 8$\times$8 blocks.  Each of these blocks are transformed using a 2D DCT transform.  Each DCT block of coefficients is further transformed using differential encoding (DC coefficients) and zig-zag encoding (AC coefficients) followed by coefficient quantisation.  These transformed and quantised coefficients are subsequently run-length entropy coded using specifically designed Huffman tables to generate the bitstream.  This process is inverted within the decoder.

As discussed above, we have investigated classification based both on the $B$ (bitstream: compressed domain) and $T$ (dequantised Transform domain) components (illustrated in figure \ref{fig:jpeg1}). These are discussed in the following subsections.

\subsection{Contributions}
The contributions of this study are as follows:
\begin{itemize}
    \item To our best knowledge, the capability of generating good classification performance directly using the bitstream (of JPEG in this case) is unique.
\item The use of the transform domain for classification has also been limited previously.  This is the first study to use the DCT transform domain for JPEG for classification.
\item The investigation of a range of network architectures for classification within the bitstream and transform domain (including an adapted ViT architecture and depth first/bit axis separable convolution layer).
\end{itemize}

\subsection{Image / Classification Datasets}
We tested both the transform domain and bitstream domain classifiers on the Caltech 101 (C101) image classification database \cite{fei2004learning}. 

The C101 database has 101 classes contained within 9,146 images They were obtained by searching class labels for 101 categories using Google's image search facility (and filtering out irrelevant results).  Each category contains between 40 to 800 images. The size
of each image is approximately 300 $\times$ 200 pixels (we resized / cropped all of the images to 256$\times$256 for consistency and utility in creating the network architectures).  The "quality" factor that was used to encode the C101 dataset was 95 and default quantisation tables were used.  

The entire C101 data of 9,146 images was divided into subsets for Training (70\%), Validation (10\%) and Testing (20\%).  The validation set is used for parameter optimisation (early stopping etc., model choice etc.) and the results are given on the testing subset.\\

\subsection{Transform Domain Classification}

A transform classification system was created based on characterising the de-quantised (rescaled) transform domain coefficients. This preserves the spatial arrangement of the underlying $8\times8$ blocks and avoids complexities associated with dependencies on quantisation tables. The utility of using the transform domain for classification is two-fold:

\begin{itemize}
    \item The DCT transform domain is a feature-rich spatial frequency representation of visual data that not only contains all the information of the image data itself but potentially offers a better representation of it for classification~\cite{hill2002wavelet}.
    \item  Full decoding is not necessary and a flexible subset of the frequency components of the 8$\times$8 DCT blocks could be used for effective classification thus reducing memory and computational requirements.
\end{itemize}

\begin{figure}[htb]
\centering
\includegraphics[width=1.03\textwidth]{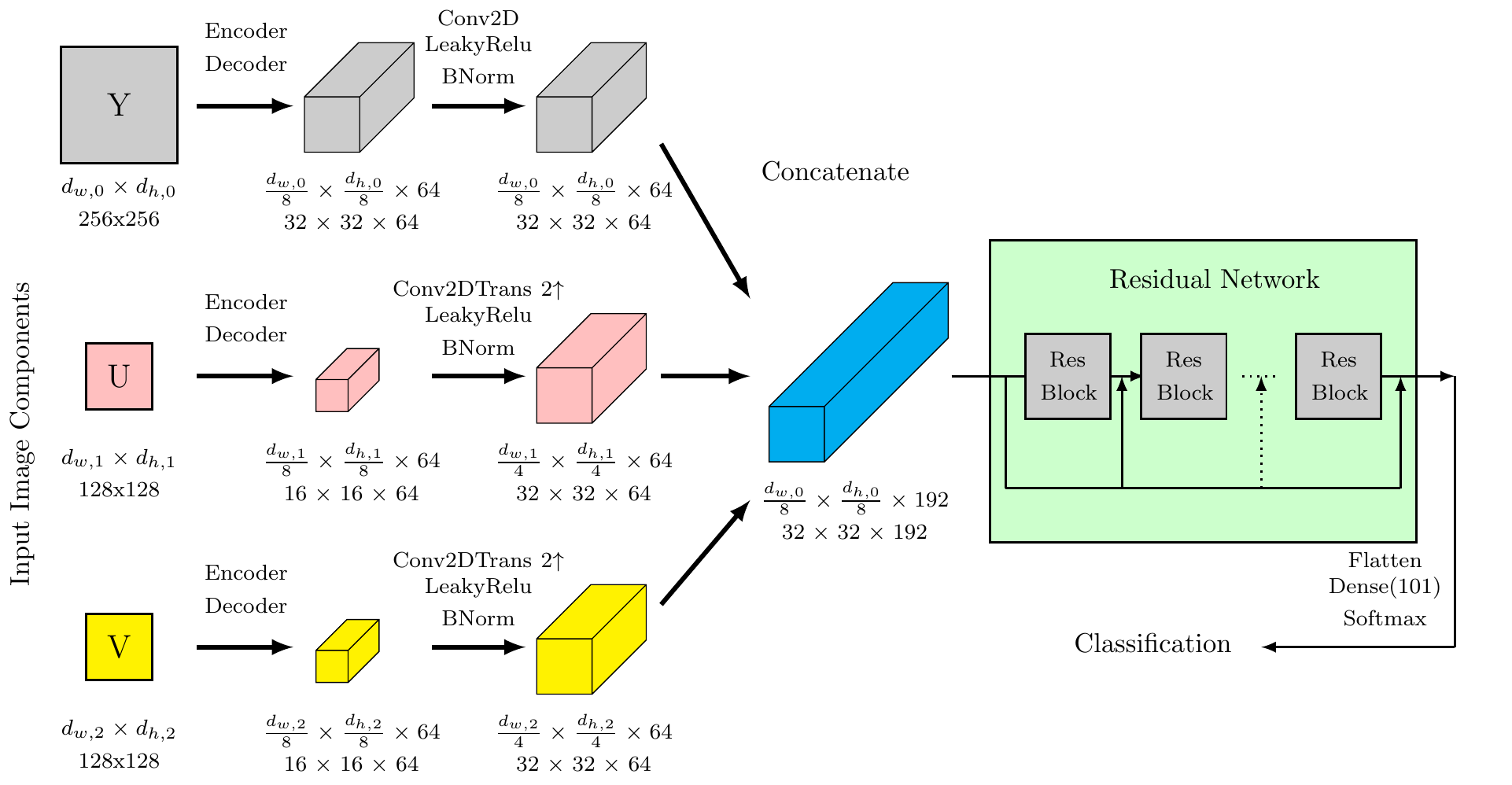}
\caption{Classification of DCT-JPEG tranform domain: C101 dataset. Input image size: $256\times256$ i.e. $d_{w,0}$=256, $d_{h,0}$=256; therefore the dimensions of the components are $d_{w,1}$=128, $d_{h,1}$=128 and $d_{w,2}$=128, $d_{h,2}$=128}

\label{fig:jpeg2}
\end{figure}

\noindent \textbf{Classification Network}\\  Figure \ref{fig:jpeg2} shows the architecture of the classification network used for classifying images within the C101 dataset.  Output 8$\times$8 DCT blocks from each image component (Y,\, U and V) are input into convolutional, concatenation and residual blocks to form a final flattened vector that uses a softmax activation function to generate a classification of each image. 
The input stages of the architecture were chosen to form equal sized tensors of the Y,U and V channels for effective combination.  The remaining layers (including the residual layers) were chosen to mirror the architecture utilised within the JPEG decoding architecture defined by Niu \cite{niu2020endtoend} (which is effectively doing the same type of analysis of JPEG transform blocks).

\begin{figure}[htb]
\center
\includegraphics[width=0.8\textwidth]{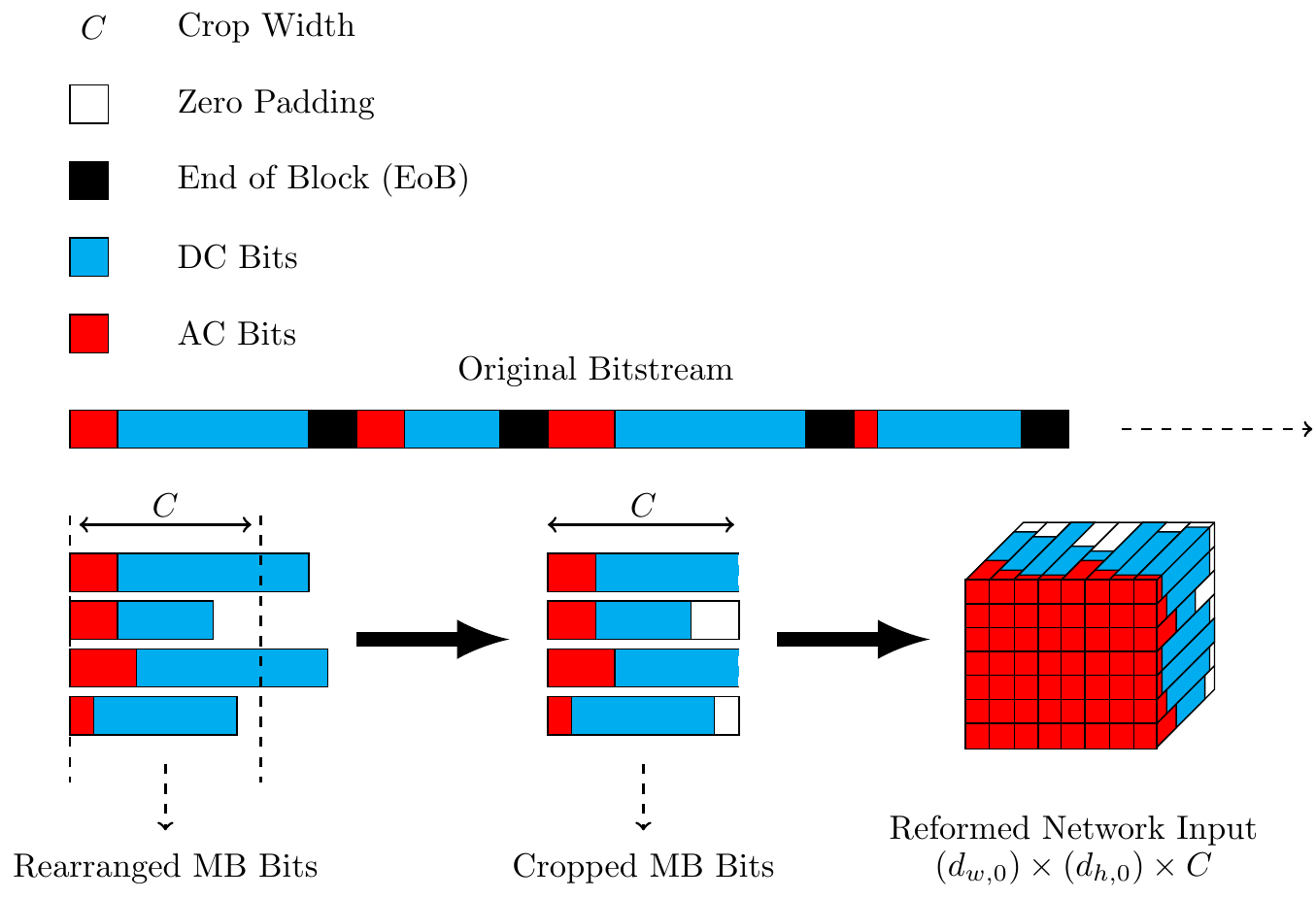}
\caption{Data preparation of DCT-JPEG bitstream domain: C101 dataset.}
\label{fig:JPEG1}
\end{figure}

\subsection{Bitstream Based Classification}
Although transform domain JPEG based classification has the advantage that no inverse transform is required in order to train and test a classifier, the alternative approach of classification using the entropy encoded (compressed) bitstream has the following advantages:
\begin{itemize}
    \item The processing and memory requirements of dealing with the data are significantly reduced; the memory requirements of using the above transform domain is the same (or larger) than the image domain  (i.e.\ the number of coefficients equals the number of pixels). 
    \item Bit errors are likely to have reduced impact on  classification performance, offering integrated error resilience.
\end{itemize}
\noindent However there are significant challenges to overcome when using this approach:
 
 \begin{itemize}
    \item We must take into account the variability of coding parameters during analysis.
    \item The entropy encoded bitstreams will be dependant on the quantisation levels and tables used to encode them.
    \item The removal of statistical redundancy through symbol encoding is likely to decrease the ability of a network to extract features for classification. 
\end{itemize}  

To overcome these problems, we simplified our system, initially using a single set of quality parameter and quantisation tables to illustrate the concept.  Obviously a practical system would need to recognise and adapt to parameter variations.  Similar approaches have been used for denoising (e.g.\ in DnCNN \cite{zhang2017beyond}). \\

\noindent \textbf{Classification Network}\\  Figure \ref{fig:jpeg2} (the classification network used for transform domain classifier) has been adapted for bitstream classification by substituting the recomposed DCT frequency cubes in figure \ref{fig:jpeg2} with the reformed network inputs, as illustrated in figure \ref{fig:JPEG1}.  These inputs are generated by segmenting all of bitstream elements of the image into 8$\times$8 macroblocks using the End of Block (EoB) markers within the bitstream.  Bits for each MB are either zero-padded or truncated according to a chosen crop width $C$. This crop width is used as the feature dimension of the input block - set to 128 bits for these experiments.\\

\noindent \textbf{Multi-Head Attention Adaptation}\\
By flattening the spatial dimensions of the Y,U and V blocks, (the stage before concatenation) the resulting 1D signal can be concatenated into a longer 1D signal.  This signal is in principle similar to the inputs into the multi-head attention stage of the so called Vision Transformers~\cite{dosovitskiy2020image}.  We therefore use two Multi-Head Attention layers feeding directly into the final dense layer (shown on the right of figure 2).  We have exhaustively considered a range of architectures and this was the best performing.

\noindent \textbf{Point Wise Separable Convolutions}\\
Conceptually, the grouping of the bitstream associated with each JPEG $8\times8$ block should be followed by a set of convolutions along the bistream dimension in order to characterise each block.  This is achieved using point wise separable convolution (instead of the default 2D convolution).
This type of convolution has been applied to both the CNN method (shown in figure 2) and the Attention method (described above).

\subsection{Results}
The compression / distortion and classification results for JPEG Bitstream and transform domain systems on the C101 dataset are given in Table \ref{tab:CCR} (according to the following methods).

\begin{description}[leftmargin=0cm, style=sameline]
\item [Method 1:] JPEG Transform domain with 2D convolutions (see figure 2).
\item [Method 2:] JPEG Bitstream domain with 2D convolutions (see figure 3).
\item [Method 3:] JPEG Bitstream domain with Point Wise Separable Convolutions (i.e.\ channel based convolutions first based on figure 3).
\item [Method 4:] JPEG Bitstream domain with Multi-Head Attention (adapted Vision Transformer).
\item [Method 5:] JPEG Bitstream domain with Multi-Head Attention (adapted Vision Transformer) together with Point Wise Separable Convolutions (i.e.\ channel based convolutions first).
\end{description}

The transform-domain classification system gives excellent performance given the difficulty of the problem (classification performance for the C101 dataset was also good (68\% accuracy)) but this does not match the current state-of-the-art (approximately 95\% accuracy).  However, our system used non-optimised components (compared to the state of the art method \cite{zhai2021scaling}.

The JPEG bitstream classification method  also produced very good performance  (~60\% accuracy for the C101 dataset).  This is quite a remarkable result given that the bitstream comprises multiple processes (including transformation, run length and entropy coding together with differential coding of the DC coefficients).  

The depth first convolution and vision transformer based methods did not perform as well as expected.  This was assumed to be caused by the capacity of simple convolution to more effectively characterise bitstream encoded macroblock data.

\begin{table}[]
\centering
\caption{Classification / Compression Results for JPEG Bitsteam and Transform Domain Systems (All JPEGs encoded with "quality" factor 95).  These results show the mean across all the testing images of the C101 dataset.}
\label{tab:CCR}
\begin{tabular}{lllll}
Database / Classifier & Accuracy & F1 & Precision & Recall \\
\hline
C101/ JPEG Transform (Method 1)        & \textbf{0.681} & 0.637 & 0.957 & 0.488   \\
C101/ JPEG Bitstream (Method 2)      & \textbf{0.595}  & 0.633 & 0.744 & 0.551   \\
C101/ JPEG Bitstream (Method 3)      & \textbf{0.578} & 0.618 & 0.748 & 0.528    \\
C101/ JPEG Bitstream (Method 4)      & \textbf{0.481} & 0.519 & 0.821 & 0.381    \\
C101/ JPEG Bitstream (Method 5)      & \textbf{0.394} & 0.369 & 0.814 & 0.240    \\
\end{tabular}
\end{table}

\section{Conclusion}
This work demonstrates the capability of using specifically designed network architectures for classification using either transform domain data or a direct analysis of the bitstream itself.  These methods were demonstrated using JPEG compressed images within the Caltech 101 (C101) database.  Although the classification results (transform domain: ~70\%, bitstream classification: ~60\%) do not approach the state of the art classification results (top-1 C101 classification results are currently ~95\% \cite{zhai2021scaling}/) this work is a clear proof of concept for these methods.  Specifically, it is clear that suitably defined data structures and network architectures are capable of characterising transformed and entropy encoded data directly on the bitstream itself.   Furthermore, the use of state of the art network components such as Multi-head attention mechanisms \cite{dosovitskiy2020image} do not appear to improve classification performance within our developed system.

Future work will include varying the amount of compression (i.e.\ Q/Qp values) to characterise the trade off between rate, distortion and classification performance.  Separate models may need to be trained for each quantisation level.  Furthermore, we will characterise the dependency of the presented technique on the analysis of End Of Block (EOB) markers.  In certain cases these are missing in JPEG bitstreams (when the zig-zag scanning proceeds to the last coefficient).  This was not found to be the case in the C101 database but could happen for lossless input data compressed using low levels of quantisation.  We will also characterise the truncation length (i.e.\ $C$ in figure 3).  In our case 128 gave the best results.  However, for images compressed to varying levels of compression the optimal value may vary.  Finally, all of the convolutional layers convert each bit to a standard precision variable within the network.  This is extremely wasteful in terms of memory and computation.  Future work will also investigate the use of binary based networks~ \cite{wang2017energy}.

\bibliographystyle{IEEEtran}%
\footnotesize\bibliography{TransDomain}

\end{document}